
\magnification=1200
\def\dsl{\raise.15ex\hbox{/}\kern-.57em\partial}
\def\Dsl{\,\raise.15ex\hbox{/}\mkern-13.5mu D} 
\def\Asl{\,\raise.15ex\hbox{/}\mkern-13.5mu A} 
\def\Bsl{\,\raise.15ex\hbox{/}\mkern-13.5mu B} 
\def\Ksl{ \overleftarrow}
\def\cc{\overrightarrow}
\font\title=cmr12 at 12pt
\footline={\ifnum\pageno=1\hfill\else\hfill\rm\folio\hfill\fi}
\baselineskip=18pt
\rightline{DFTUZ 92/13}
\rightline{IPNO-TH 92/82}
\rightline{September 1992}
\vskip 1.0cm
\centerline{\title SECOND ORDER FORMALISM}
\centerline{{\title FOR FERMIONS}\footnote{$^\dagger$}{Work partially supported
by CICYT (Proyecto AEN 90-0030).}}
\vskip 2.0cm
\centerline{\bf J. L. Cort\'es$^1$, J. Gamboa$^2$ and
L. Vel\'azquez$^1$}
\centerline{\it $^1$Departamento de F\'\i sica Te\'orica, Universidad de
Zaragoza,}  \centerline{\it 50009 Zaragoza, Spain.}
\centerline{\it $^2$Divisi\'on de Physique The\'orique, Institut de Physique
Nucleaire\footnote{$^\ddagger$}{\it Unite de Recherche des Universit\'es
Paris 11 et Paris 6 associe au CNRS},}
\centerline{\it F-91406 Cedex, Orsay, France.}

\vskip 1.0cm
{\bf Abstract}. A new formulation of fermions based on a second order action
is proposed. An analysis of the U(1) anomaly allows us to test the validity
of the formalism at the quantum level. This formulation gives a new perspective
to the introduction of parity non invariant interactions.
\vfill \eject

In order to avoid the difficulties of the Klein-Gordon equation which results
from a correspondence principle applied to the relativistic case, Dirac {\bf
[1]} was lead to introduce a first order relativistic equation which describes
spin-${1\over 2}$ particles. The associated first order lagrangian, in
constrast to the second order bosonic lagrangian, is the root of many of the
characteristic properties of half integer spin particles. In particular the
peculiarities of the quantization of a fermionic field and the correspondence
of physical degrees of freedom with field components, the axial anomaly in even
dimensions and the species doubling which appears when a fermionic field is
formulated on the lattice, are some examples of features which are related to
the use of a first order action.

Since the original difficulties in the Klein-Gordon equation are still present
in the Dirac action, a natural question is whether it is possible to find a
path integral formulation of fermions based on a second order action. It is not
the first time that this idea has been considered {\bf [2]}. The study of the
way how all the properties which seem to be a direct consequence of a  first
order action can be understood in the second order formulation together with a
first look at the possible advantages of using a second order fermionic action,
is the main subject of this letter.

 A formulation based on the use of two component spinors
instead of Dirac spinors and its non-trivial consequences on the introduction
of parity violating
interactions as well as its  possible application to the understanding of
the weak decays, was considered long ago {\bf [3]}\footnote {$^\ast$}{ Some
related aspects of the canonical quantization of the Feynman and Gell-Mann
formulation and some variations of it were considered in {\bf [4]}.}. For a
recent attempt to use these ideas to identify a lattice formulation of chiral
gauge theories see {\bf [5]}.

At present the fashionable way to introduce parity violating interactions
through a generalization to the chiral case of the gauge principle which allows
to  understand the strong and electromagnetic interactions, has lead to a
formulation of chiral gauge theories were many open problems remain to be
solved. It is then probably interesting to keep an open mind on this problem
and to try new ways to introduce parity violating interactions.

The second order formalism of fermions presented in this work, which is based
on
the identification of decoupled field components in the Dirac action, provides
a
framework were the introduction of parity non invariant interactions presents
new aspects which have not an analog in the parity conserving case.

The method we follow to obtain a second order formulation for fermions is to
translate at the level of the path integral formulation the derivation of a
second order equation for spinors.  One way to do that is to rewrite the Dirac
equation as a set of two first order coupled equations. The starting point in
the path integral formulation is the (euclidean) first order gauge invariant
action
$$ S^{(1)} =  \int d^Dx \,\,\biggl[\,\, m {\bar \psi}\psi + {1\over 2}{\bar
\psi} ( {\cc {\Dsl}} -  {\Ksl{\Dsl}}) \psi \biggr] \eqno(1)$$
where
$$ {\cc {D}}_\mu  \psi = \left( \partial_\mu \psi - ie A_\mu \psi\right),
$$  $$ {\bar \psi} {\Ksl {D}}_\mu = \left( \partial_\mu {\bar \psi} + ie A_\mu
{\bar \psi}\right). \eqno(2) $$ We assume an even space-time dimension D in
order to decompose the Dirac field $\psi$ $$\psi = \pmatrix{\psi_L \cr 0\cr} +
\pmatrix{0 \cr \psi_R\cr} =  \biggl({1 + \gamma_{D+1}\over 2}\biggr) \psi +
\biggl({1 - \gamma_{D+1}\over 2}\biggr) \psi , $$
$${\bar \psi} = ({\bar \psi_R}\,\,\, 0) + (0 \,\,\, {\bar \psi_L}) =
 {\bar \psi}\biggl({1 + \gamma_{D+1}\over 2}\biggr) +
 {\bar \psi}\biggl({1 - \gamma_{D+1}\over 2}\biggr), \eqno(3) $$
where ${\gamma_{D+1}}^2 = 1, \,\,\, \{ \gamma_{D+1}, \gamma_\mu\} = 0$ and then
$$ {\bar \psi}\Dsl \psi = {\bar \psi}_L {\Dsl }_{L} \psi_L +
{\bar \psi}_R {\Dsl }_{R} \psi_R. \eqno(4)$$

If one introduces the variables
$$\chi_L = {1\over \sqrt{m}} \psi_L, \,\,\,\,\, \chi_R = \sqrt{m} ( \psi_R  +
{\cc {\Dsl}_{L}\over m}\psi_L )$$
$${\bar \chi_R} = {1\over \sqrt{m}} {\bar \psi_R} , \,\,\,\,\, {\bar \chi_L} =
\sqrt{m} ( {\bar \psi_L}  -  {\bar \psi}_R { {\Ksl{\Dsl}}_{R}\over m} ),
\eqno(5)$$
then the first order action takes the form
$$\eqalignno{ S^{(1)} =  \int d^Dx \biggl[&\,\, m^2 {\bar \chi}_R{ \chi}_L -
{1\over 2}  {\bar \chi}_R ( {\cc {\Dsl }}_{R} {\cc {\Dsl}}_{L} +
{\Ksl {\Dsl }}_{R}{\Ksl {\Dsl }}_{L})\chi_L
+ {\bar \chi_L}\chi_R  \cr
& +{1\over 2} {\bar \chi}_R (
{\cc {\Dsl }}_{R} + {\Ksl {\Dsl }}_{R})\chi_R  -   {1\over 2} {\bar \chi}_L
 ({\Ksl {\Dsl }}_{L} + {\cc {\Dsl }}_{L}) {\chi_{L}}  \biggr], &(6) \cr}$$
where the last two terms which mix the two components reduce to a total
derivative. The variables  $\chi_R,{\bar \chi}_L$ are auxiliary fields, they
are
decoupled from the gauge field and do not propagate. Then one can consider
$$S^{(2)} = \int d^Dx \,\,\biggl[\,\, m^2 {\bar \chi}_R{ \chi}_L +
  {\bar \chi}_R ( {\Ksl {\Dsl }_{R}}{\cc {\Dsl }}_{L} )\chi_L
\biggr],\eqno(7)$$
as the action for the second order fermionic formulation with all the dynamics
concentrated on the anticonmutating fields ${\bar
\chi}_R, \chi_L$. One could anticipate from a naive counting of degrees of
freedom, that a translation to a second order formulation must be acompanied by
a decoupling of half of the original fields.

One important point to remark is that a mass term is essential in order to go
from the first order action to the second order formulation. The massless case
is special from this point of view as well as the case of a chiral gauge
theory where a mass term is not allowed by gauge invariance.

The previous steps going from the Dirac fermionic action to the second order
action $S^{(2)}$ in (7) guarantee the equivalence of both formulations at the
classical level. For instance, it is straightforward to reconstruct the free
Dirac propagator by combining the propagators of the fermionic fields
$\chi, {\bar \chi}$ calculated with the action $S^{(2)}$ in the free case
$( D_\mu \to \partial_\mu)$ and the relation (5) between the original Dirac
fields $\psi$ and the fields $\chi, {\bar \chi}$. What remains to be studied in
order to establish the validity of the second order formulation is to see that
the quantum fluctuations do not destroy the equivalence with the Dirac first
order
 action.

In this sense one crucial point is to see whether the axial anomaly {\bf [6]}
is
reproduced in the present formulation. In order to study this question we will
concentrate for simplicity on the abelian two dimensional case. Using the Weyl
representation for the two dimensional euclidean gamma matrices
$$\gamma_1 = \sigma_2, \,\,\,\,\, \gamma_2 = -\sigma_1$$
$$\gamma_5 = -i\gamma_1 \gamma_2 = \sigma_3, \eqno(8)$$
and introducing for any vector the chiral components
$$a_{\pm} =  a_1 \pm ia_2, \eqno(9)$$
then one has in this case
$$\Dsl_{R} = -iD_- $$
$$\Dsl_{L} = iD_+, \eqno(10)$$ and the second order lagrangian is
given by
$${\cal L}^{(2)} = \{  m^2 {\bar \chi}_R{ \chi}_L -
  {{\bar \chi}_R} D_- D_+ \chi_L \} + \lq\lq
total\,\,\,derivatives". \eqno(11)$$
In order to study how the anomaly is reproduced in the second order
formulation let us first rewrite the standard Fujikawa derivation of the
anomaly in terms of the variables $\chi , {\bar \chi} $ in (5). This will
help us to understand the way the anomaly appears in the second order
formalism. In the Fujikawa  method {\bf [7]} the anomaly is understood as a
non invariance of the fermionic measure under the axial transformation. The
fermionic measure is defined in terms of the coefficients of the expansion of
the fermionic field in eigenfuntions of the Dirac operator $${\cal D}\psi {\cal
D}{\bar \psi} = \prod_n da_n d {\bar a_n}, \eqno(12)$$ with  $$\psi = \sum_n
a_n
{\varphi}_n$$ $${\bar \psi}= \sum_n {\bar a_n} {\varphi}^+_n, \eqno(13)$$ and
$$\Dsl \varphi_n = \lambda_n \varphi_n. \eqno(14)$$

If one makes use for the eigenfuntions $\varphi_n$ of the spinorial
decomposition
in (3)
$${\varphi}_n = \pmatrix{\phi_n \cr 0\cr} + \pmatrix{0 \cr
\eta_n\cr},\eqno(15)$$  and taking into account the spinorial structure of the
two dimensional Dirac operator in (10) one has
$$D_- D_+ \phi_n = \lambda_n^2 \phi_n$$
$$D_+ D_- \eta_n = \lambda_n^2 \eta_n. \eqno(16)$$

Since
$$\Dsl \gamma_5 \varphi_n = -\lambda_n \gamma_5 \varphi_n$$
we can consider $\lambda_n > 0$ (we do not consider the effect of zero modes
in this discussion) and there is a one to one correspondence between the
measure of the Dirac field and a measure of the components $\psi_L, \psi_R$
through the expansion in eigenfuntions of the operators
$ D_- D_+$ and $D_+ D_-$
$${\cal D}\psi_L {\cal D} \psi_R = \prod_n da_n^L d a^R_n,
\eqno(17)$$
with
$$ \psi_L = \sum_n a^L_n  \phi_n, $$
$$ \psi_R = \sum_n a^R_n \eta_n. \eqno(18)$$

If this expansion is plugged in (5) then it results into
$$ \chi_L = \sum_n b^L_n  \phi_n, $$
$$ \chi_R = \sum_n b^R_n \eta_n. \eqno(19)$$
which leads to a fermionic measure
$${\cal D}\chi_L {\cal D} \chi_R = \prod_n db_n^L db^R_n,
 \eqno(20)$$
and a related factor for ${\bar \chi}$. Then the fermionic measure for the
variables $\chi_L, \chi_R$ is defined in terms of the eigenfuntions of the
operators $D_-D_+$ and $D_+D_-$ respectively.

Once the measure has been identified then one can repeat the standard Fujikawa
evaluation of the anomaly. First one sees the effect over the measure,through
the change of the coefficients in the expansion of a chiral transformation
$\psi^{\prime} (x) = e^{i \alpha (x) \gamma_5} \psi (x)$
$$\prod_n  db^{'L}_n db^{'R}_n  = \left(1 - i \int d^2x \alpha (x)
A(x) \right) \,\,\prod_n  db^L_n db^R_n , \eqno(21)$$
where $A(x) = \lim_{M \to \infty} \biggl[ A^L(x) - A^R(x) \biggr] $ and
$$A^L(x) =  \sum_n \phi_n^+ (x) \exp [ -{D_-D_+\over
M^2}]\phi_n (x),\eqno(22) $$
 $$A^R(x) = \sum_n \eta_n^+ (x) \exp [ -{D_-D_+\over
M^2}]\eta_n (x),\eqno(23) $$
$A^L, A^R$ being the constributions to the anomaly from the two components
$\chi_L, \chi_R$. The origin of the anomaly from this point of view is the
difference $(D_+D_- \not= D_-D_+)$ between the operators involved in the two
factors of the fermionic measure. The last step is to use the basis
independence of the trace involved in $A^L , A^R$ to replace the eigenfuntions
$\phi_n, \eta_n$ by plane waves which allow a straightforward generalization of
the Fujikawa's calculation. Making use of the identities
$$D_- D_+ = D_\mu D_\mu + i F_{12}, \eqno(24)$$
$$D_+ D_- = D_\mu D_\mu - i F_{12}, \eqno(25)$$
one obtains
$$A(x) = -{1\over 2\pi}F_{12}, \eqno(26)$$
which can be combined with a similar contribution from ${\bar \chi}$ to
reproduce the standard form of the axial 2-dimensional anomaly. One remark to
be made from this derivation of the anomaly is that the variable $\chi_R$
plays an important role in this evaluation; even though the $\chi_R$ component
is  completely decoupled at the level of the action it still envolves a gauge
field  dependence in the definition of the measure which shows up in the axial
anomaly.

Note that the axial $U(1)$ global transformation of the original Dirac field
$$\psi^{\prime}_L = e^{i\alpha} \psi_L , \,\,\,\,\,
\psi^{\prime}_R  = e^{-i\alpha} \psi_R,$$
translates into
$$\eqalignno{&\chi^\prime_L = e^{i\alpha} \chi_L ,\cr &
\chi^\prime_R =
e^{-i\alpha} \chi_R +( e^{i\alpha} -  e^{-i\alpha}) \Dsl_{L} \chi_L,
&(27)\cr}$$
which involves the decoupled component $\chi_R$ in a non-trivial way.

Now, one can ask if the second order formulation, where $\chi_R $ is absent,
will be able to reproduce completely the anomaly. We will answer this question
by a direct calculation of the anomaly in the second order formulation. Our
derivation is based on the effective gauge field action generated by the
fermion field fluctuations, which formally can be written as  $$\Gamma = -\ln
\det \biggl[ {(\Dsl_{R} \Dsl_{L} - m^2) \over (\partial_-  \partial_+ - m^2)}
\biggr]. \eqno(28)$$  In order to study the anomaly one has to introduce an
external field $A^5_\mu$ which acts as a source for the axial current.  This is
done by replacing the Dirac operator $\Dsl$ by $${\cal \Dsl} = \Dsl -i
\gamma_\mu
\gamma_5 A^5_\mu, \eqno(29)$$ and the two dimensional anomaly can be identified
from the term linear in the fields $A_\mu$ and $A^5_\mu$ in the
expansion of the effective action  $\Gamma (A, A^5)$  $$\Gamma_2 (A,A^5) = e
\int {d^2k\over {(2\pi) }^2}  A^5_\mu (-k) \Pi^5_{\mu\nu} (k) A_\nu (k).
\eqno(30)$$ When we calculate the effective action for the first order
formulation the standard result for  the U(1) anomaly follows directly from the
rotational and gauge invariance of the Dirac action (1). Since the term ${\bar
\chi_L}\chi_R $ neglected when going to the second order formalism respects
these symmetries one can expect that the same result for the U(1) anomaly will
be obtained when the second order action (7) is used. Let us see explicitly the
derivation of this result from a direct calculation of the axial polarization
tensor $\Pi^5_{\mu\nu}.$ When the effective action in (28),with $D$ replaced by
${\cal D}$ ,is expanded  in powers of $A_\mu$ and $A^5_\mu$ then one obtains
$$\eqalignno{& \Pi^5_{+-} = 4 \int {d^2l\over {(2\pi) }^2} \biggl[ P(l) -  {(l
+
{k\over 2})}_-  {(l + {k\over 2})}_+ P(l + {k\over 2}) P(l - {k\over 2})
\biggr], &(31a)  \cr & \Pi^5_{++} = -4 \int {d^2l\over {(2\pi)}^2}  {(l +
{k\over 2})}_+  {(l - {k\over 2})}_+ P(l + {k\over 2}) P(l - {k\over 2}),
&(31b)  \cr & \Pi^5_{--} = 4 \int {d^2l\over {(2\pi)}^2}  {(l + {k\over 2})}_-
{(l - {k\over 2})}_- P(l + {k\over 2}) P(l - {k\over 2}), &(31c)  \cr &
\Pi^5_{-+} = -4 \int {d^2l\over {(2\pi)}^2} \biggl[ P(l) -  {(l - {k\over
2})}_+  {(l - {k\over 2})}_- P(l + {k\over 2}) P(l - {k\over 2}) \biggr],
&(31d)
\cr}$$  where $$P(q) = {[ q^2 + m^2]}^{-1}, \eqno(32)$$  is the bosonic free
propagator.

The integrals in (31) are
logarithmically divergent and an ultraviolet regularization is required in
order to evaluate the effective action. It is convenient to make a
decomposition

$$\Pi^5_{\mu\nu} = {\bar \Pi^5_{\mu\nu}} + {\tilde \Pi^5_{\mu\nu} }.
\eqno(33)$$
The first term is given by the axial polarization tensor with a
substraction of the integrand at zero external momentum in order to have a
convergent integral. All the regularization dependence will be
concentrated on the momentum independent contribution
${\tilde \Pi^5_{\mu\nu}}$.

The regularization independent contribution can be written as
$$\eqalignno{&{\bar \Pi}_{++} = -4 I_{++}, \,\,\,\,\, {\bar \Pi}_{--} = 4
I_{--},
\cr & {\bar \Pi}_{+-} = -{\bar \Pi}_{-+} = 4 m^2 B, &(34) \cr}$$
where
$$\eqalignno{&I_{\mu\nu} = \int {d^2l\over {(2\pi )}^2} \biggl\{
 {(l + {k\over 2})}_\mu  {(l - {k\over 2})}_\nu P(l + {k\over 2}) P(l - {k\over
2}) - l_\mu l_\nu P^2 (l) \biggr\}, &(35)
\cr & B = \int {d^2l\over {(2\pi )}^2} \biggl\{
 P(l + {k\over 2}) P(l - {k\over 2}) - P^2 (l)\biggr\}, &(36) \cr }$$
and by covariance arguments $$I_{\mu\nu} = C\,{k_\mu \, k_\nu \over k^2} + D\,
g_{\mu\nu} , \eqno(37)$$ where $m^2B,\,C,$ and $D$ will be functions of
${k^2\over m^2}$.

If one uses a gauge invariant regularization of the effective action then
$$k_\nu \Pi^5_{\mu\nu} = k_\nu {\bar \Pi}^5_{\mu\nu} +
k_\nu {\tilde \Pi}^5_{\mu\nu} = 0, \eqno(38)$$
and this is enough to fix the regularization dependent component ${\tilde \Pi}$
$$\eqalignno{&{\tilde \Pi}_{++} = {\tilde \Pi}_{--} = 0,
 \cr & {\tilde \Pi}_{+-} = -{\tilde \Pi}_{-+} = -4(m^2B - C). &(39) \cr} $$

When this result is combined with ${\bar \Pi}$ in (34) one finds
$$\Pi^5_{\mu\nu} = 4iC
\epsilon_{\rho\nu} {k_\mu k_\rho\over k^2}, \eqno(40)$$
and the anomaly can be read from
$$k_\mu \Pi^5_{\mu\nu} = 4iC k_\rho \epsilon_{\rho \nu}. \eqno(41)$$

In fact, since the axial polarization tensor is related to the effective
action by (30), then one has in the limit $ m \to 0 $
$${\partial \over \partial x_\mu} {\delta \Gamma (A, A^5)\over \delta A^5_\mu
(x)} \biggr\vert_{A^5 = 0} = 2ie\,C(m=0) \epsilon_{\rho\nu}F_{\rho\nu} =
- {ie \over 2\pi } \epsilon_{\rho\nu}F_{\rho\nu} ,
\eqno(42)$$ which is the standard result for the anomaly of the axial current
as
obtained  from the first order Dirac action.
Then one can rederive the anomaly from the effective action of the second order
formulation, which makes manifest the decoupling at the quantum level of the
auxiliary field $\chi_R$. This is in contrast with the derivation of the
anomaly from the fermionic measure where $\chi_R$ was an essential ingredient.
In fact the current one is considering in the second order formulation is
given by
$$ J^{(2)}_{\mu 5} (x) = {\delta S^{(2)}\over \delta A^5_{\mu} (x)} =-i \biggl[
{\bar \chi}_R {\Ksl {\Dsl }}_{R} {(\gamma_\mu )}_{L} \chi_L  +
{\bar \chi }_R {(\gamma_\mu)}_{R} {\cc {\Dsl }}_{L} \chi_L \biggr],
\eqno(43)
$$
which is the Noether current associated to the
transformation
$$\chi^{\prime}_L = e^{i\alpha} \chi_L , \,\,\,\,\,
{\bar \chi^{\prime}}_R  = {\bar \chi}_R e^{i\alpha} . \eqno(44)$$

Then the perturbative calculation based on the efective action leads to
$$\partial_\mu < J^{(2)}_{\mu 5} > = \partial_\mu {\delta \Gamma \over
\delta A^5_\mu (x)} = <\delta S^{(2)} > + 2A^L \eqno(45)$$
where the first term on the right side  takes into in account the variation of
the second order action under the global transformation (44) and the second
term
$2A^L$ is the contribution to the Ward identity from the measure which in the
second order formalism involves $\chi_L, {\bar \chi}_R$ exclusively.

The
contribution to the anomaly from the decoupled variables is included in the
second order formulation in the variation of the action under a chiral
transformation of the dynamical variables. This is the way the anomaly is
reproduced in this context.

A particular case where the previous analysis of the
anomaly can be tested, after a regularization is introduced in order to define
the theory, is the lattice formulation of the second order formalism {\bf [8]}.

It is natural to expect that the previous analysis can be directly extended to
the non-abelian case as well as to $D>2$ and that the equivalence at the
quantum
level of the first and second order formulations is also valid in the general
case.

The symmetric way the two chiralities are treated in the usual Dirac lagrangian
is lost by the identification of the decoupled variables  required in order to
go to the second order formulation. Then one can ask how the parity
invariance of the Dirac action is reflected in the second order action. If one
eliminates the auxiliary field $\chi_R$ in the parity transformation law of the
Dirac field one is lead to consider
$$\chi^{\prime}_L (x) = {\tilde {\Dsl}_{L} \over m } \chi_L ({\tilde x}),
\eqno(46)$$
where ${\tilde x}$ is the parity transformed of $x$ and ${\tilde {\Dsl} }$
is the corresponding covariant derivative
$${\tilde D_\mu } = {\partial \over \partial {\tilde x}^\mu } -
ie  A_\mu ({\tilde x}) . \eqno(47)$$

When the transfomation (46) is applied to the second order lagrangian one gets
$${{\cal L}^{(2)}}^\prime \left( x \right)  = {\cal L}^{(2)} \left( {\tilde x}
\right) + \delta {\cal L}^{(2)} \left( {\tilde x} \right) . \eqno(48)$$
If one uses the identities
$${{\Dsl}^\prime_{L}} = {\tilde {\Dsl}}_{R},\,\,\,\,\,
{{\Dsl}^\prime_{R}} = {\tilde {\Dsl}}_{L}, \eqno(49)$$
where ${D_\mu }^\prime = {\partial \over \partial  x^\mu } -
ie  A^\prime_\mu (x) ,$ then the variation of the lagrangian under parity is
given by
$$ \delta {\cal L}^{(2)} = {\displaystyle{\bar \chi}}_R  {{\Ksl {\Dsl
}_{R}}{\Ksl {\Dsl }}_{L}\over m}   \,. \,  {{\cc {\Dsl }_{R}}{\cc {\Dsl
}}_{L}\over m} \chi_L
 - m^2 {\bar \chi}_R{\displaystyle{ \chi}}_L \eqno(50)$$
which vanishes if $\chi_L ({\bar \chi}_R)$ is a solution of the classical
equations of motion. Note that the elimination of the variables $\chi_R, {\bar
\chi_L}$ when  going to the second order formulation can be understood as a
consequence of the equations of motion; then it is  natural for the equation
of motion of the dynamical degrees of freedom to be required in order to
identify the invariance under parity of the second order action \footnote
{$^\ast$}{ In fact if one applies twice the transformation (46) then
$  \chi_L^{\prime \prime} ={1 \over m^2} {\cc {\Dsl }_{R}}{\cc {\Dsl
}}_{L} \chi_L $ and therefore this transformation is involutive only on the
classical configurations.}.

It is the requirement of parity invariance what prevents from adding new
renormalizable interactions directly in the second order formalism. A
possible contact interaction
$({\bar \chi}_R \chi_L) ({\bar \chi}_R \chi_L)$ should be accompanied by the
corresponding non-renormalizable terms involving derivatives, as obtained from
the interaction  written in terms of the Dirac field
$({\bar \psi} \psi) ({\bar \psi} \psi )$.

The derivation of the second order lagrangian from the Dirac lagrangian
requires to consider a massive fermion (see eq. (5)). But nothing prevents to
consider directly
$${\cal L}^{(2)} = {\bar \chi}_R {\Ksl {\Dsl}}_{R} {\cc {\Dsl}}_{L} \chi_L,
\eqno(51)$$
as a candidate for a massless second order lagrangian and to generalize the
gauge parity conserving interaction to the chiral case by considering
$$ {\cc {D}}_\mu  \chi_L = \left( \partial_\mu \chi_L - ie A_\mu \chi_L\right),
$$  $$ {\bar \chi}_R {\Ksl {D}}_\mu = \left( \partial_\mu {\bar \chi}_R +
i{\bar
e} A_\mu  {\bar \chi}_R\right), \eqno(52) $$
with $e \ne {\bar e}$ and the obvious non-abelian generalization.

Note that the second order lagrangian (51) is not invariant under the local
transfomation
$$\eqalignno{& \chi^{\prime}_L = e^{ie\alpha (x)} \chi_L (x), \,\,\,\,\,
{\bar \chi}^\prime_R  = {\bar \chi}_R (x) e^{-i{\bar e}\alpha (x)} ,
\cr & A^\prime_\mu (x) = A_\mu (x) + \partial_\mu \alpha (x),  &(53) \cr}$$
for the same reason the second order action was not invariant under a global
$U(1)$ transformation (44) in the parity conserving case.

If one considers an anomaly free fermion content then there will be a double
cancellation of contributions, first at the level of the fermionic measure and
second for the expectation value of the variation of the action. The
possibility
to formulate a chiral gauge theory along this lines and the possibility to
include new renormalizable interactions in the second order formulation, like
four fermion  interactions involving the dynamical fields ${\chi }_L ,\,
{{\bar \chi }}_R $ deserves further investigation.

To summarize, a second order formulation based on the
identification of a  combination of fermionic field components with no dynamics
in any even dimensions of spacetime, has been proposed as a way to study a
gauge
invariant parity conserving theory. The possibility to apply this formalism to
the case of a chiral gauge theory has been pointed out with the perspectives
that it can open on the dynamics of these theories.

We would like to thank M. Asorey and H.B. Nielsen by useful
discussions.
\vskip 0.25cm
\centerline{\bf References}

\item{\bf [1]} P.A.M. Dirac, Proc. of the Roy. Soc. {\bf117}(1928)610.
\item{\bf [2]} R.P. Feynman and M. Gell-Mann, Phys. Rev. {\bf 109}(1958)193.
\item{\bf [3]} T.W.B. Kibble and J.C. Polkhinhorne, Nuov. Cim. {\bf
8}(1958)1001.
\item{\bf [4]} F. Palumbo, Nuov. Cim. {\bf
104A}(1991)1851 .
\item{\bf [5]} T. Banks and A. Casher, Nucl. Phys. {\bf 169B}(1980)103;
A.C. Longhitano and B. Svetitsky, Phys Lett. {\bf 126B}(1983)259.
\item{\bf [6]} S.L. Adler, Phys. Rev. {\bf 177}(1969)2426; J.S. Bell and R.
Jackiw,  Nuov. Cim. {\bf 60}(1969)47; W.A. Bardeen, Phys. Rev. {\bf
184}(1969)1848.
\item{\bf [7]}  K. Fujikawa, Phys. Rev. {\bf 21D}(1980)2848;
Phys. Rev. {\bf 22D}(1980)1499.
\item{\bf [8]} J.L. Cort\'es, J. Gamboa and L. Vel\'azquez, {\it
Second Order  Formalism for Fermions: Anomalies and Species Doubling on the
Lattice}, Zaragoza-Orsay Preprint.
\end